%% file: abun_hds_100giants.tex
\documentclass{pasj00}

\draft

\begin{document}
\SetRunningHead{Author(s) in page-head}{Running Head}

\title{Stellar Parameters and Chemical Abundances of G Giants}

\author{Liang \textsc{Wang}}
\affil{
Key Laboratory of Optical Astronomy,
National Astronomical Observatories,
Chinese Academy of Sciences,
A20, Datun Road, Chaoyang District, Beijing 100012, China
}
\affil{
Graduate University of the Chinese Academy of Sciences,
19A Yuquan Road, Shijingshan District,
Beijing 100049, China
}
\email{wangliang@nao.cas.cn}

\author{Yujuan \textsc{Liu}}
\affil{
Key Laboratory of Optical Astronomy,
National Astronomical Observatories,
Chinese Academy of Sciences,
A20, Datun Road, Chaoyang District, Beijing 100012, China
}\email{lyj@nao.cas.cn}

\author{Gang \textsc{Zhao}}
\affil{
Key Laboratory of Optical Astronomy,
National Astronomical Observatories,
Chinese Academy of Sciences,
A20, Datun Road, Chaoyang District, Beijing 100012, China
}\email{gzhao@nao.cas.cn}

\and
\author{Bun'ei \textsc{Sato}}
\affil{
Tokyo Institute of Technology,
2-12-1 Ookayama, Meguro-ku,
Tokyo 152-8550, Japan
}\email{sato.b.aa@m.titech.ac.jp}


\maketitle

\begin{abstract}
We present basic stellar parameters of 99 late-type G giants
    based on high resolution spectra obtained by the High Dispersion Spectrograph attached to Subaru Telescope.
These stars are targets of a Doppler survey program searching for extra-solar planets among evolved stars,
    with a metallicity of $-0.8<\mathrm{[Fe/H]}<+0.2$.
We also derived their abundances of 15 chemical elements, including
    four $\alpha$-elements (Mg, Si, Ca, Ti),
    three odd-Z light elements (Al, K, Sc),
    four iron peak elements (V, Cr, Fe, Ni),
    and four neutron-capture elements (Y, Ba, La, Eu).
Kinematic properties reveal that most of the program stars belong to the thin disk.
\end{abstract}

\section{Introduction}

Since the first extra-solar planet around main sequence star, 51 Peg b was discovered in 1995 \citep{Mayor1995},
    more than 500 planets have been revealed in solar neighbourhood \footnote{see http://exoplanets.org/},
    providing clues on planet formation process around different types of stars.
However, only a small fraction of known planet-host stars are intermediate-mass ($1.5\sim5\mathrm{M}_{\odot}$).
Properties of planets around these stars are of great importance to constrain planet formation theories (e.g. \cite{IdaLin2004}).
One way to search for extra-solar planets around intermediate-mass stars is monitoring radial velocities of
    late-type G, K giants.
Their spectra have many sharp absorption lines, and therefore suitable
    for getting precise radial velocity data by modern Doppler techniques (e.g. \cite{Butler1996}).

The East Asian Planets Search Network (EAPSNet, see \cite{EAPSNet})
    which was started in 2005 has established an international consortium
    among Japanese, Korean and Chinese researchers.
More than 500 late-type G, K giants are being monitored with 2-m class telescopes among three countries.
Stellar parameters and detailed elemental abundances of some of the targets have been analysed by \citet{Takeda2008} and \citet{Liu2010}.
As an extension of EAPSNet, the Subaru Planets Search Program started monitoring radial velocity
    variations of about 300 late-type G, K giants in 2006.
By the end of June 2010, two planets (HD 145457b and HD 180314b) have been discovered in this program
    \citep{Sato2010}.
While basic parameters and chemical abundance analysis have not been conducted for this sample.

In this paper we present the first results of stellar parameters and chemical abundances for 99 giants based on
    high resolution spectra obtained with Subaru Telescope.
In Section 2, we describe the observation and data reduction method.
Section 3 gives the derived stellar parameters, including the comparison of different methods.
Chemical abundances and kinematic properties are presented in Section 4 and 5, respectively.
In the last section, conclusions are presented.

\section{Observation and data reduction}

Targets of Subaru Planet Search Program were selected from {\em Hipparcos Catalogue} \citep{hip} by the following criteria,
\begin{enumerate}
\item a colour index of $0.6 \lesssim B-V \lesssim 1.0$ and an absolute magnitude of $-3 \lesssim M_\mathrm{V} \lesssim 2.5$,
    corresponding to the range of evolved, late-type G giants.
\item a magnitude of $6.5 < V < 7.0$, which enables follow-up observations using 2m-class telescopes.
\item a declination of $\delta > -25^\circ$ so as to be observable in Japan, Korea and China.
\end{enumerate}
First, each target was observed with the High Dispersion Spectrograph (HDS; \cite{HDS}) equipped to Subaru Telescope
    three times with an interval of about 1.5 months,
    for the purpose of quickly identifying stars with large radial velocity variations.
Then the most likely candidates with sub-stellar companions were repeatedly observed
    with the 1.88m telescope at Okayama Astrophysical Observatory (OAO, Japan),
    1.8m telescope at Bohyunsan Optical Astronomy Observatory (BOAO, Korea),
    and 2.16m telescope at Xinglong Station (National Astronomical Observatories, China).

HDS has several setups of gratings,
    and each has different spectrum format and wavelength coverage.
In the first three runs of this program in 2006,
    we used StdI2b twice and StdI2a once.
The spectra taken with StdI2a setup with a wavelength coverage of 4900$\sim$7600 \AA\ 
    are suitable for chemical abundance analysis.
Here we present the first results of these 99 stars whose spectra were obtained in July, 2006,
    with a resolving power of $R\sim60,000$ and a typical signal-to-noise ratio (SNR) of 150-230/pixel.
The bias is determined and subtracted using the over-scan region on CCD,
    with an IRAF script provided by HDS web page \footnote{see http://www.naoj.org/Observing/Instruments/HDS/}.
Then, the reduction of data follows the standard routines.
The MIDAS/ECHELLE package is used for order definition,
    flat-fielding, background subtraction, 1-D spectra extraction, and wavelength calibration.
The radial velocity shift is corrected by fitting profiles of about 80 pre-selected lines with intermediate strength.
Although the Doppler shift correction is made on the un-normalized spectra,
    the spectra lines are narrow (typically $\sim\mathrm{0.3\ \AA}$)
    so that the continumm around such a line could be seen as a constant,
    and nearly do not affect the measured central wavelength.
The uncertainties of radial velocity values are about $0.3\sim0.4\,\mathrm{km\,s^{-1}}$
    using this method (see table \ref{tab:kinetics}).
After that, the one dimension spectra are normalized by a continuum fit by a cubic spline function
    with a smooth parameter.
The spline interpolation is determined by a set of continuum windows selected from a typical giant spectrum,
    usually has a density of 10-15 points per order.

For the purpose of precise radial velocity measurement,
    the spectra were taken with an iodine vapor cell inserted into the light path of the spectrograph,
    hence the stellar spectral lines located in the wavelength of 5000$\sim$6300 \AA\ region
    were heavily mixed with absorption features in the $\mathrm{I_2}$ spectrum.
We only use the red part of the spectra with wavelength longer than 6350 \AA\ to analyse the chemical abundances.
Equivalent widths are measured by fitting the line profiles with a Gaussian function for the weak lines.
While in strong unblended lines, broad damping wings contribute significantly to the equivalent width.
The direct integration is used for such kind of lines alternatively.

\section{Stellar parameters}

\subsection{Effective temperature}
\begin{figure}
  \begin{center}
     \FigureFile(70mm, 70mm){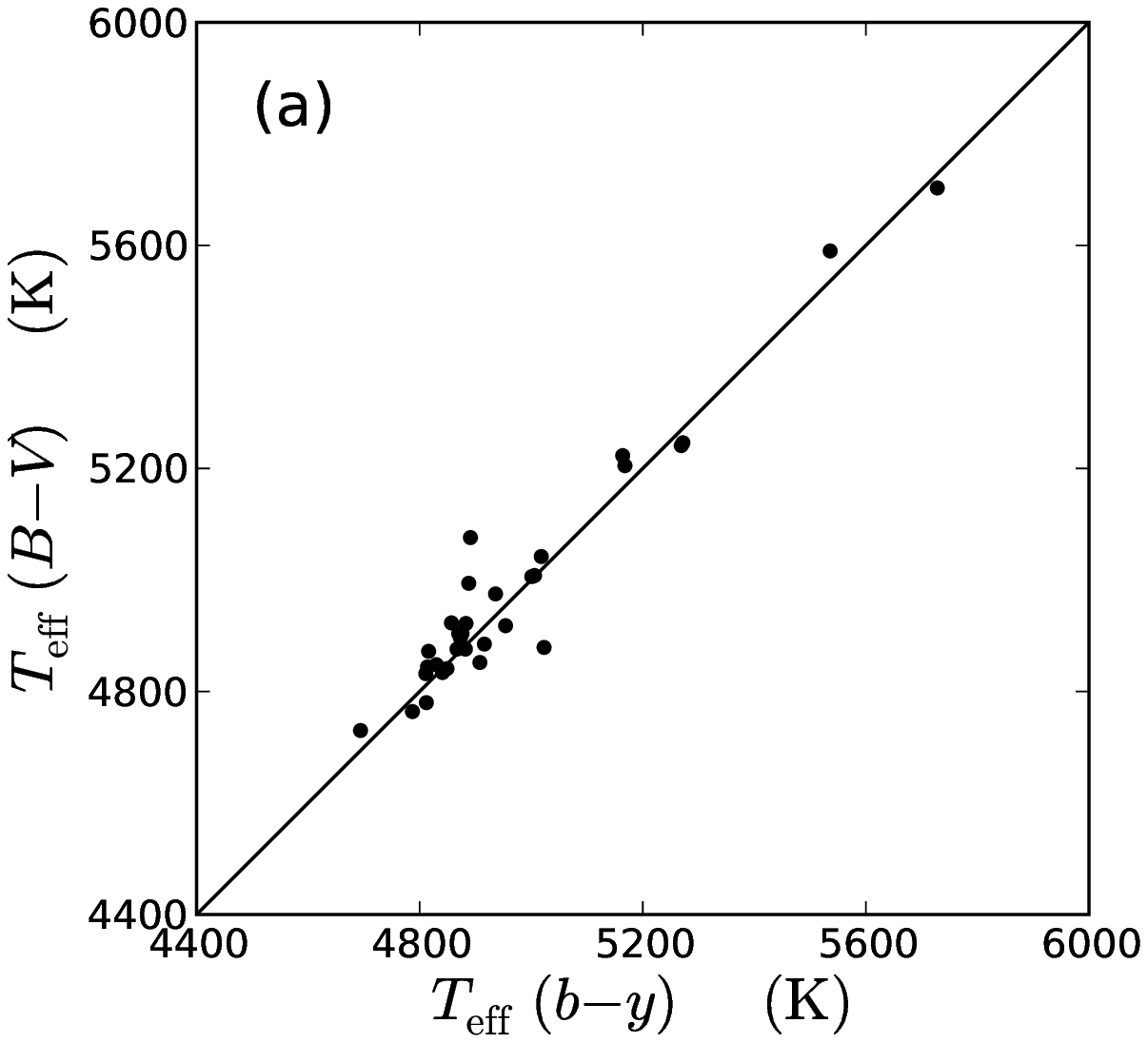}
     \FigureFile(70mm, 70mm){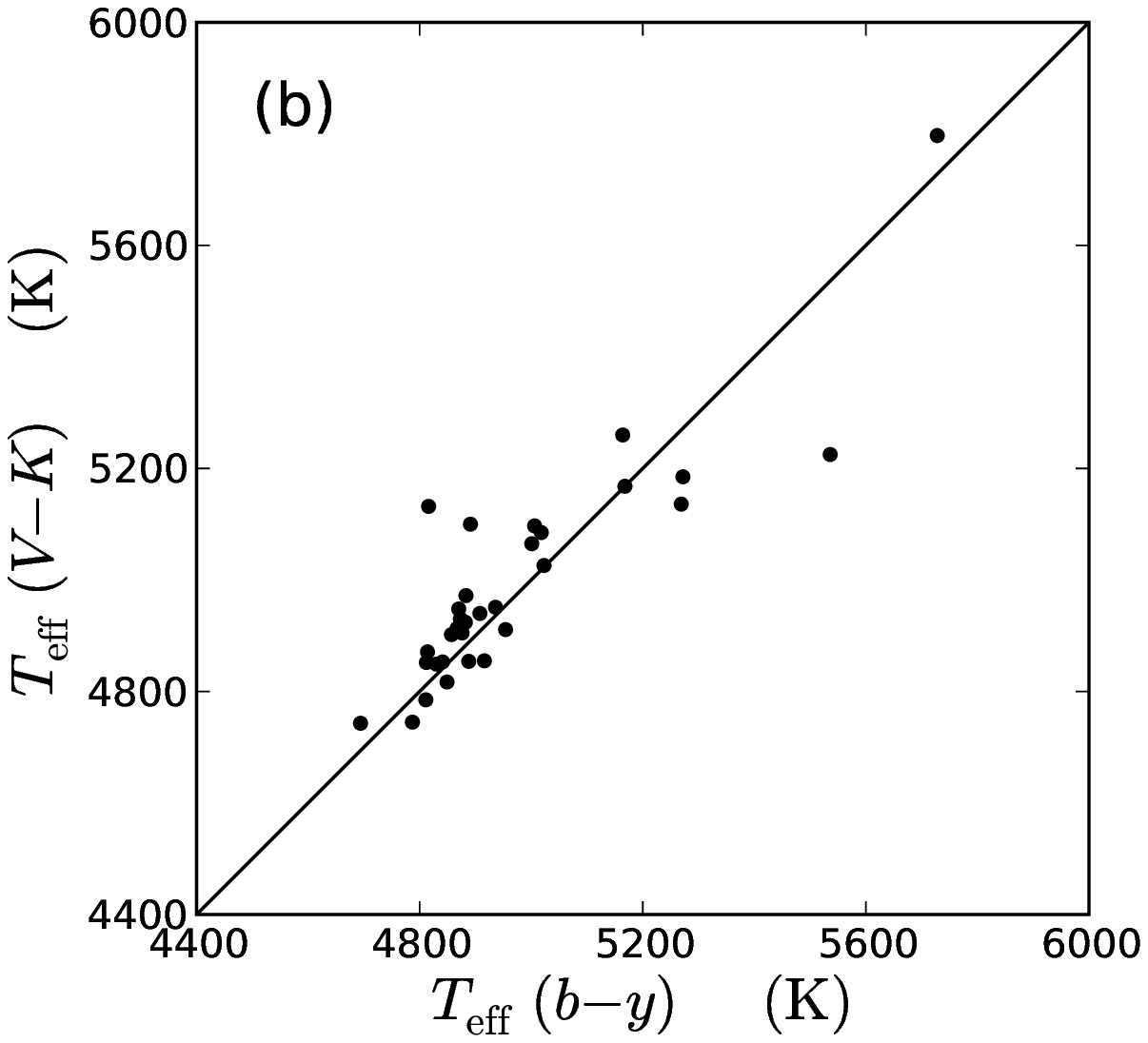}
  \end{center}
  \caption{Comparison of effective temperatures derived from different colour indices.
         (a): $T_\mathrm{eff}$ derived from $B-V$ versus that from $b-y$.
         (b): $T_\mathrm{eff}$ derived from $V-K$ versus that from $b-y$.}
  \label{fig:comp-teff1}
\end{figure}

Based on the empirical calibration relations given by Alonso et al. (1999, 2001),
    the effective temperatures ($T_\mathrm{eff}$) of our program stars are determined
    using the $B-V$, $V-K$, and $b-y$ photometric data.
The $B-V$ colour indices in Johnson system are obtained by
    $B-V=0.850(B_{\mathrm{T}}-V_{\mathrm{T}})$, as described by \citet{hip1200},
    where $B_\mathrm{T}$ and $V_\mathrm{T}$ are Tycho magnitudes in {\em Hipparcos Catalogue} \citep{hip}.
The $K$ magnitudes in $V-K$ are obtained by converting the $K_\mathrm{s}$ magnitude
    in Two Micron All Sky Survey (2MASS) to $K$ magnitude using the calibration relation given by \citet{Ramirez2004},
    and $V$ magnitudes are taken from {\em Hipparcos Catalogue}.
There are 32 stars of our sample also have $b-y$ data in catalogue given by \citet{Hauck1998}.
As a result, we derive the $T_\mathrm{eff}$ from $b-y$ besides $V-K$ and $B-V$, and compare the results with those from $B-V$ and $V-K$
    in figure \ref{fig:comp-teff1}.

The color excess $E(B-V)_{\mathrm{A}}$ is calculated according to \citet{Schlegel1998},
    with a slightly revision described by \citet{Arce1999} for those values
    larger than 0.15 mag.
Then, the $E(B-V)$ value for each star is calculated as
    $$E(B-V)=[1-\exp{(-|D\sin{b}|/125)}]E(B-V)_{\mathrm{A}}$$
    where $D$ is the distance of the star and $b$ is the galactic latitude.
Finally, we adopt $E(V-K)=2.948E(B-V)$ as colour excess for $V-K$ \citep{Schlegel1998},
    and $E(b-y)=0.741E(B-V)$ for $b-y$ \citep{Crawford1976}.

The mean difference $\langle T_\mathrm{eff}(B-V)-T_\mathrm{eff}(b-y)\rangle$ is $16\pm56\,\mathrm{K}$,
    while the mean difference $\langle T_\mathrm{eff}(V-K)-T_\mathrm{eff}(b-y)\rangle$ is $24\pm100\,\mathrm{K}$.
According to \citet{Alonso1999}, the uncertainty of the effective temperature of giants derived from $V-K$ is estimated
    to be $\pm 40\,\mathrm{K}$, which is more accurate than that is derived from either $B-V$ ($\pm 96\,\mathrm{K}$)
    or $b-y$ ($\pm 70\,\mathrm{K}$).
However, most stars in our sample are too bright in $K_\mathrm{s}$ band ($<4^\mathrm{m}.5$)
    to get accurate photometry,
    therefore the effective temperatures derived from $V-K$ may not be reliable.
Table \ref{tab:comp_teff} list the effective temperatures derived from three different 
    colour indices for the 32 stars.

{\footnotesize
\begin{table}[htbp]
\caption{Comparison of effective temperatures derived from three different colour indices.}
\label{tab:comp_teff}
\begin{tabular}{c|ccc||c|ccc||c|ccc}
\hline
HD &  $T_\mathrm{eff}^{b-y}$ & $T_\mathrm{eff}^{B-V}$ & $T_\mathrm{eff}^{V-K}$ &
HD &  $T_\mathrm{eff}^{b-y}$ & $T_\mathrm{eff}^{B-V}$ & $T_\mathrm{eff}^{V-K}$ &
HD &  $T_\mathrm{eff}^{b-y}$ & $T_\mathrm{eff}^{B-V}$ & $T_\mathrm{eff}^{V-K}$ \\
\hline
\input{table.teff_comp.dat}
\hline
\end{tabular}
\end{table}
}

\begin{figure}
  \begin{center}
     \FigureFile(70mm, 70mm){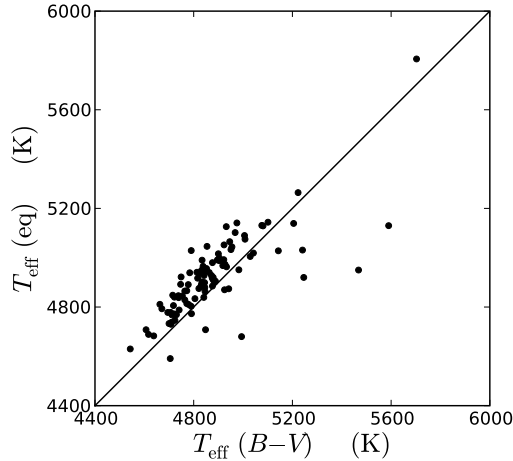}
  \end{center}
  \caption{Comparison of effective temperatures derived from $B-V$
           and excitation equilibrium methods.}
  \label{fig:comp-teff2}
\end{figure}

Another method is the so-called ``excitation equilibrium method'', which forces the derived iron abundances
    given by different FeI lines to be independent from their excitation potentials of lower states ($\chi_\mathrm{low}$).
After adjusting effective temperatures, the slopes on $\log A - $$\chi_\mathrm{low}$ diagrams for all stars 
    in our sample are smaller than 0.002 dex/eV,
and the mean difference $\langle T_\mathrm{eff}(\mathrm{eq})-T_\mathrm{eff}(B-V)\rangle$ is $44\pm117\,\mathrm{K}$.
The effective temperatures derived from this method and those from $B-V$
    are compared in figure \ref{fig:comp-teff2}.

\subsection{Surface gravity}
Surface gravity ($\log g$) is determined by
\begin{equation}
\label{math:logg}
\log{g} = \log{g_\odot} +
          \log\left(\frac{M}{M_\odot}\right) + 
          4\log\left(\frac{T_\mathrm{eff}}{T_{\mathrm{eff},\odot}}\right) +
          0.4(M_{\mathrm{bol}}-M_{\mathrm{bol},\odot})
\end{equation}

where $M$ is the stellar mass, and $M_\mathrm{bol}$ is the bolometric magnitude defined as
\begin{equation}
\label{math:Mbol}
    M_{\mathrm{bol}}=V_\mathrm{mag}+BC-5\log{d}+5-A_{\mathrm{V}}
\end{equation}
where $V_\mathrm{mag}$, $BC$, $d$, and $A_{\mathrm{V}}$ represent
    the apparent magnitude, bolometric correction, distance,
    and interstellar extinction, respectively.

The bolometric corrections ($BC$) are calculated based on estimated effective temperatures and metallicties,
    as given in \citet{Alonso1999}.
The stellar mass is estimated by an interpolation of Yale-Yonsei stellar evolution tracks
    (\cite{YYtrack2004}, \cite{YYtrack2003})
    which is based on a new convective core overshoot scheme.
Yale-Yonsei evolution tracks can be used for stars from the stage of the pre-main-sequence
    to the helium-core flash.
For comparison, we also estimated the masses using the evolution tracks of \citet{Girardi2000},
    in which the helium-core ingnition phase is included.
By assuming all the stars are corresponding to the post-RGB phase,
    we find the difference between the two sets of stellar mass is less than $0.3\,M_\odot$.
While about half of our program stars present an uncertainty of parallax up to 15\%.
According to equation (\ref{math:logg}) and (\ref{math:Mbol}), this will cause the error of $\sim0.13$ dex in $\log g$,
    and about a factor of 1.3 in stellar mass.
The interstellar extinctions are adopted by $A_\mathrm{v}=3.1E(B-V)$,
    and {\em Hipparcos} parallaxes are used to determine the absolute magnitudes ($M_\mathrm{V}$).

\begin{figure}
  \begin{center}
     \FigureFile(70mm, 70mm){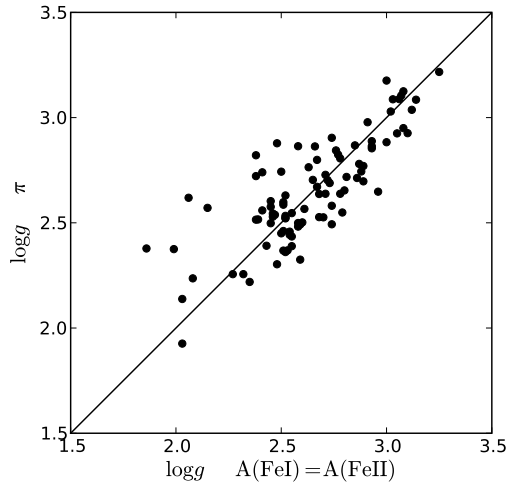}
  \end{center}
  \caption{Comparison of surface gravities obtained by {\em Hipparcos} parallaxes (Y axis)
           and ionization balance of FeI and FeII lines (X axis). }
  \label{fig:comp_logg}
\end{figure}

We also adopt another method which is independent from parallax data to determine the surface gravities.
As the FeII lines are more sensitive to surface gravities in cool stars than FeI lines,
    $\log g$ can be determined by forcing the abundances of FeII lines to be the
    same value as those given by FeI lines.
Despite several uncertainties such as non-local thermodynamic equilibrium (NLTE) effects and
    very few Fe II lines, it is still meaningful to compare different sets of $\log g$ values.
As shown in figure \ref{fig:comp_logg}, the scatter tends to increase with decreasing surface gravities.

\subsection{Atomic data}
The $\log gf$ values for selected spectral lines are taken from various literatures, and listed in table 2.
For Fe I lines, we use Oxford oscillator strength data (Blackwell 1982a,b, 1986).
For the four heavy elements (Y, Ba, La, Eu), the $\log gf$ values are taken from \citet{Hannaford1982},
    \citet{Weise1980}, \citet{Luck1991}, and \citet{Biemont1982}, respectively.
For other elements, we use the $\log gf$ values given by \citet{Chen00}.
Although the atomic data are compiled from different references,
    it does not affect the results systematically.
Moreover, the same program and values of oscillator strength are also adopted by
    measuring the EWs in Solar Atlas \citep{Kurucz1984sun} to determine the solar abundances,
    so the final results are the differential values relative to the Sun.

{\footnotesize
\begin{table}[htbp]
\caption{
    Atomic line data.
    }
\label{tab:line}
\begin{tabular}{ccccc}
\hline
$\lambda_\mathrm{air}$ (\AA) & Element & $\chi_\mathrm{low}$ &
$\log gf$ & $E_\gamma$ \\
\hline
 6327.604 & Ni I &1.68 &-3.150 &2.5\\
 6335.337 & Fe I &2.20 &-2.177 &1.2\\
 6336.830 & Fe I &3.69 &-0.856 &1.4\\
 6344.155 & Fe I &2.43 &-2.923 &1.3\\
   ...    &  ... & ... & ...   &...\\
\hline
\end{tabular}
\end{table}
}

\subsection{Metallicity and microturbulence velocity}
For most of our program stars, the initial metallicity values are set to [Fe/H]=0.0.
For those stars that have been studied, the initial values are taken from previous literatures
    (\cite{silva2006}, \cite{Mishenina2006}, \cite{Luck2007}, and \cite{Takeda2008}).
The abundances of chemical elements are determined based on the model atmospheres interpolated by a plane-parallel,
    homogeneous and local thermodynamic equilibrium (LTE) model grid by \citet{KuruczCD13}.
Chemical abundances are calculated with ABONTEST8 program supplied by Dr. P. Magain (Liege, Belgium).
It calculates the theoretical equivalent widths from the atmospheric model,
    and matches them with observed values.
Several broadening mechanism have been taken into account,
    including natural broadening, thermal broadening, van der Waals damping broadening, and microturbulent broadening.
The microturbulence $\xi_\mathrm{t}$ is determined by forcing the iron abundance values given by different Fe I lines
    to be independent from their EWs and hence with zero slope on $\log A$-$EW$ diagram.
Only those Fe I lines with 10 m\AA\ $< EW <$ 110 m\AA\ are used.

\subsection{Summary}
The derived physical parameters of our program stars are summarized in table \ref{tab:para_BV} and \ref{tab:para_eq}.
In table \ref{tab:para_BV}, effective temperatures are derived from $B-V$ and 
    and surface gravities are derived from {\em Hipparcos} parallaxes.
While in table \ref{tab:para_eq}, effective temperatures and surface gravities are
    derived from excitation equilibrium and and ionization balance, respectively.
Although two different sets of stellar parameters are given, we finally adopt
    those listed in table \ref{tab:para_BV} in the following analysis.

{\footnotesize
\begin{table}[htbp]
\caption{Physical parameters of program stars.
    Effective temperatures are derived from $B-V$ and calibration relation of \citet{Alonso1999}.
    Surface gravities are derived from {\em Hipparcos} parallaxes.}
\label{tab:para_BV}
\begin{tabular}{c|ccccccccccc}
\hline
HD&$T_\mathrm{eff}$ (K)&$\mathrm{[Fe/H]}$&$\log g$&$\xi_\mathrm{t} (\mathrm{km\,s^{-1}})$&
$B-V$ & $E(B-V)$ & $M_\mathrm{v}$ & $BC$ & $M/M_\odot$ &
$\log age$ & $\log(L/L_\odot)$ \\
\hline
 100055 & 4900 & -0.16 & 2.780 & 1.79 & 0.929 & 0.008 &  0.720 & -0.296 & 2.29 & 8.90 & 1.77 \\
 101853 & 4834 & -0.21 & 2.549 & 1.76 & 0.955 & 0.019 &  0.098 & -0.322 & 2.58 & 8.75 & 2.00 \\
 103690 & 4730 & -0.32 & 2.499 & 1.61 & 0.996 & 0.010 &  0.560 & -0.366 & 1.70 & 9.24 & 1.84 \\
 105475 & 4764 & -0.10 & 2.638 & 1.46 & 1.007 & 0.016 &  0.345 & -0.351 & 2.74 & 8.69 & 1.94 \\
  ...   &  ... &  ...  &  ...  & ...  &  ...  &  ...  &   ...  &   ...  & ...  & ...  & ...  \\

\hline
\end{tabular}
\end{table}
}

{\footnotesize
\begin{table}[htbp]
\caption{Atmospheric parameters of program stars.
    Effective temperatures are derived from excitation equilibrium method.
    Surface gravities are derived from ionization balance method.}
\label{tab:para_eq}
\begin{tabular}{c|cccc}
\hline
  HD   & $T_\mathrm{eff}$ (K) & $\mathrm{[Fe/H]}$ & $\log g$ & $\xi_\mathrm{t} (\mathrm{km\,s^{-1}})$ \\
\hline
 100055 & 5016 & -0.07 & 2.87 &  1.76 \\
 101853 & 4990 & -0.09 & 2.79 &  1.76 \\
 103690 & 4769 & -0.29 & 2.58 &  1.60 \\
 105475 & 4865 & -0.03 & 2.71 &  1.44 \\
  ...   &  ... &  ...  & ...  &  ...  \\
\hline
\end{tabular}
\end{table}
}

\begin{figure}
  \begin{center}
    \FigureFile(70mm, 70mm){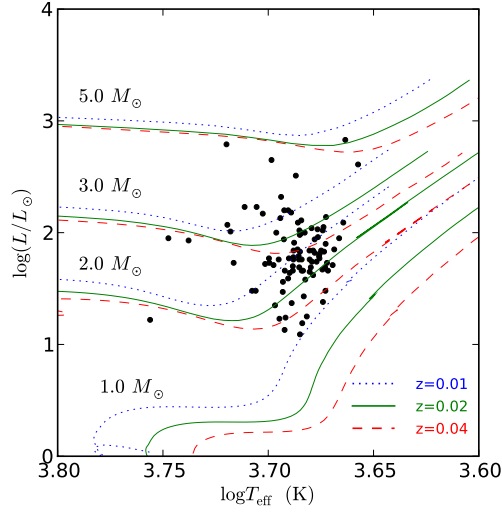}
  \end{center}
  \caption{Luminosity versus effective temperature of our program stars,
           together with the Yale-Yonsei evolution tracks \citep{YYtrack2004} of stars
           with different masses ($1\,M_\odot$, $2\,M_\odot$, $3\,M_\odot$, and $5\,M_\odot$,)
           and metallicity (z=0.04, 0.02, and 0.01, corresponding to [Fe/H]
           =+0.3, 0.0, and -0.3, respectively).}
  \label{fig:hrd}
\end{figure}

\begin{figure}
  \begin{center}
    \FigureFile(70mm, 50mm){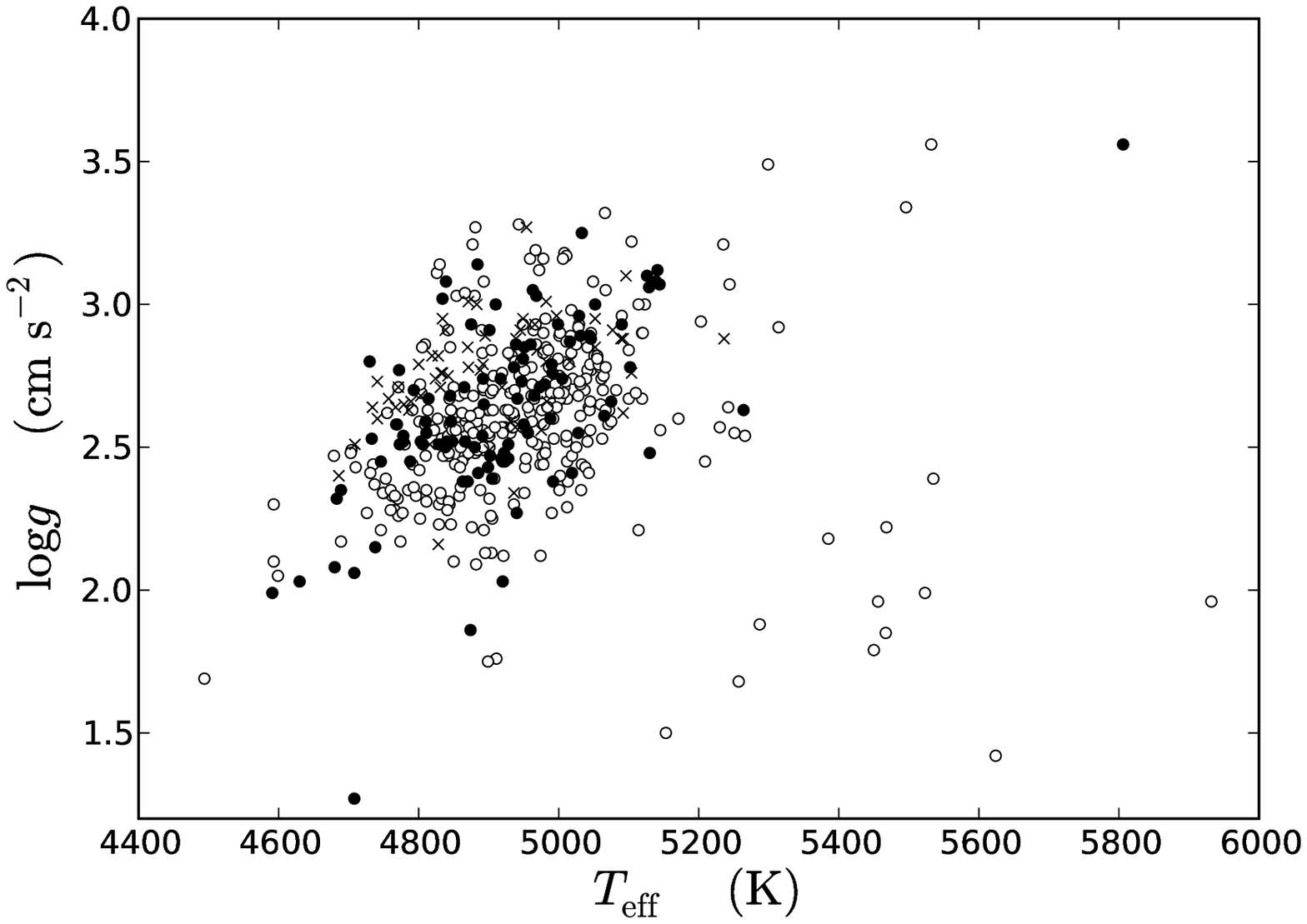}
    \FigureFile(70mm, 50mm){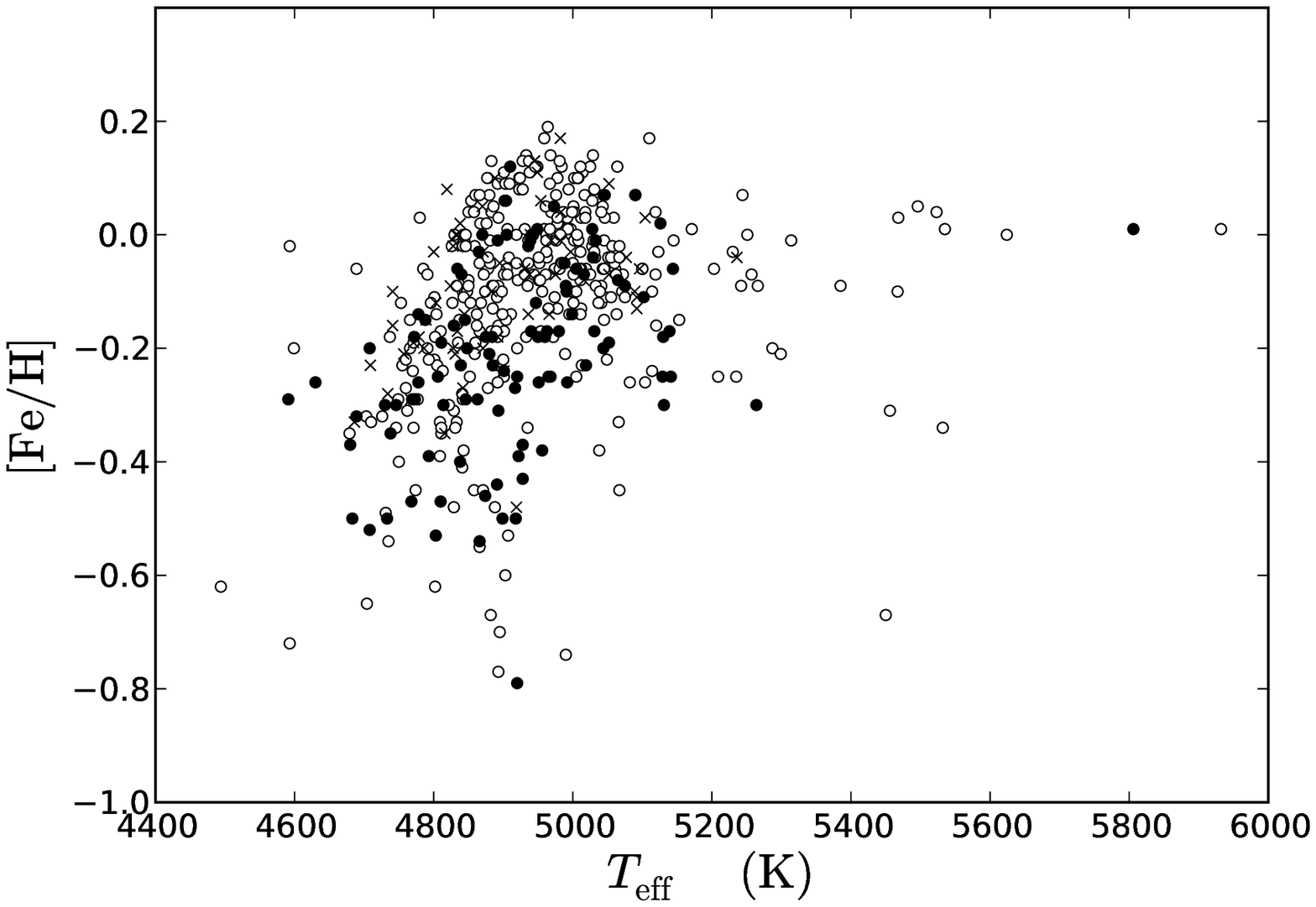}
    \FigureFile(70mm, 50mm){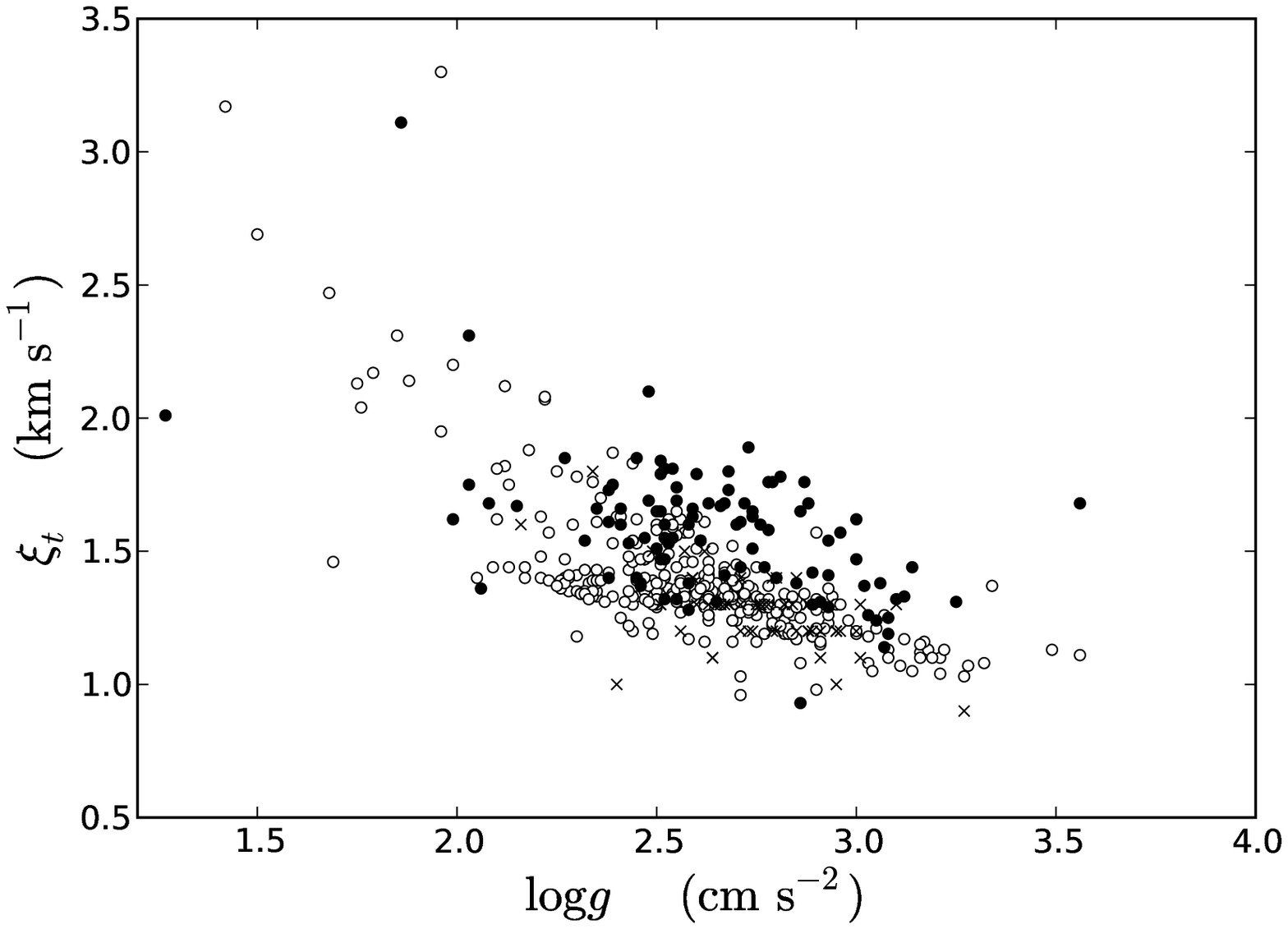}
  \end{center}
  \caption{
           Relations of atmospheric parameters
           ($\log g$ vs. $T_\mathrm{eff}$,
           [Fe/H] vs. $T_\mathrm{eff}$,
           $\xi_t$ vs. $\log g$)
           derived from $B-V$.
           Filled circles: this study,
           open circles: \citet{Takeda2008},
           and crosses: \citet{Liu2010}.
           }
  \label{fig:relation}
\end{figure}

In figure \ref{fig:hrd} $\log(L/L_\odot)$ versus $T_\mathrm{eff}$ is plotted for our program stars.
It is shown that most of our samples have luminosities $1.5<\log L/L_\odot<2.0$,
    and effective temperatures $3.67<\log T/T_\mathrm{eff}<3.70$.
In figure \ref{fig:relation}, the derived atmospheric parameters of our stars (filled circles)
    are compared with those derived by \citet{Takeda2008} (open circles) and \citet{Liu2010} (crosses).
The effective temperature ($T_\mathrm{eff}$) tends to increase with increasing surface gravity ($\log g$),
    and the microturbulent velocity ($\xi_\mathrm{t}$) tends to decrease with higher surface gravity ($\log g$).

\subsection{Consistency check with literatures}
\begin{figure}
  \begin{center}
     \FigureFile(70mm,70mm){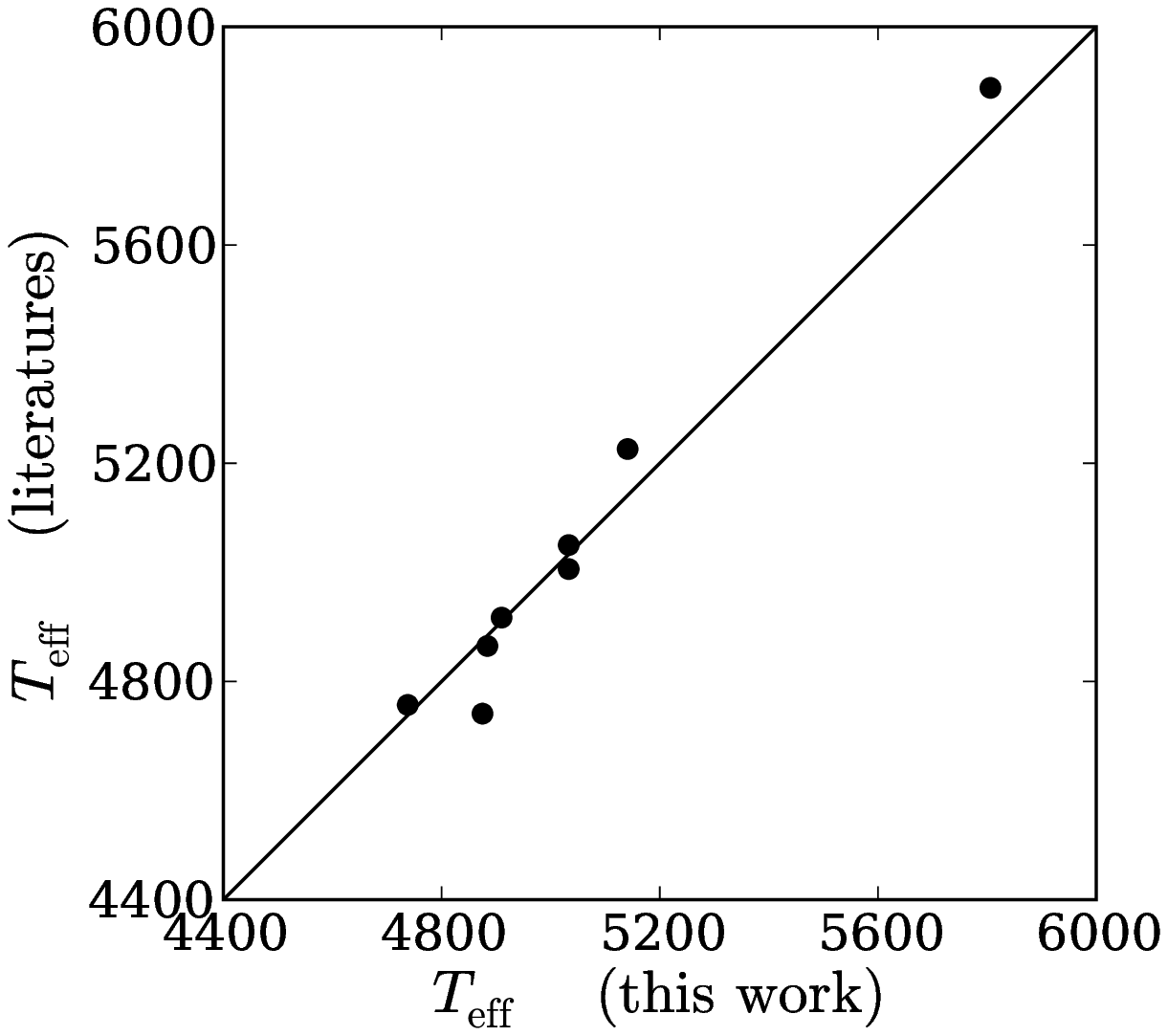}
     \FigureFile(70mm,70mm){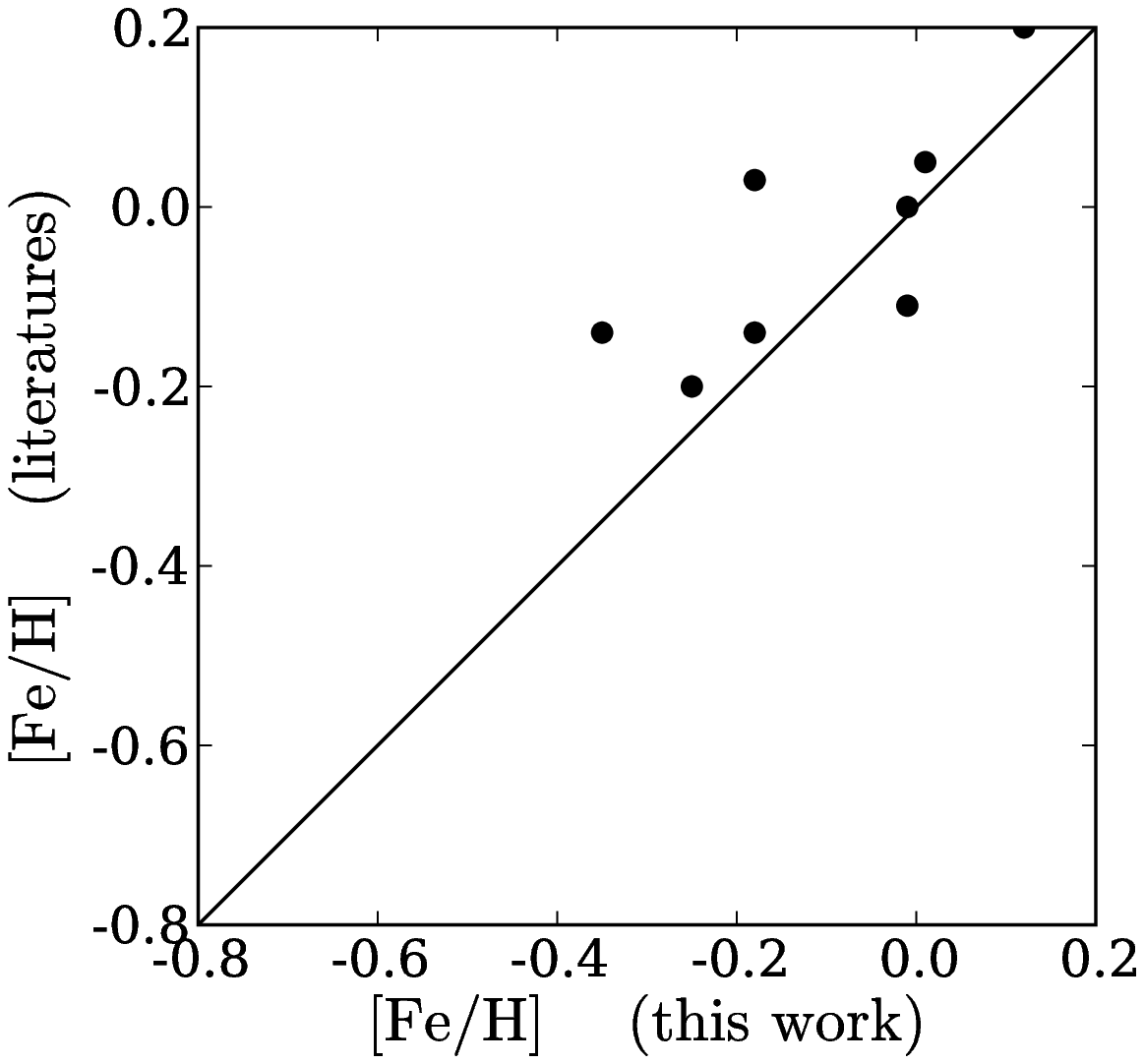}
     \FigureFile(70mm,70mm){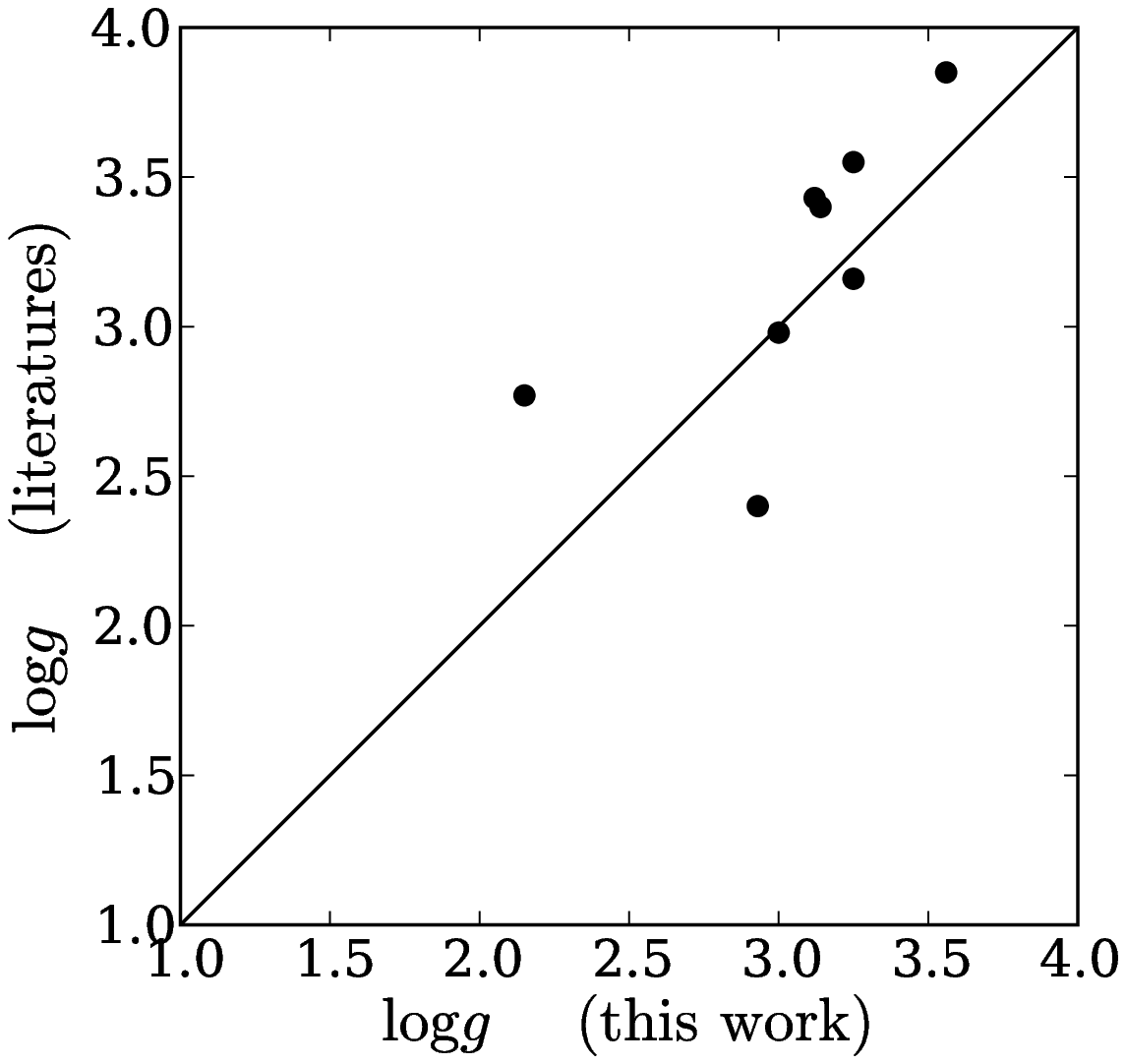}
  \end{center}
  \caption{
           Comparison of atmospheric parameters of common stars
           in this study and other literatures.
           Upper panel: effective temperature,
           middle panel: metallicity,
           bottom panel: surface gravity.
           }
  \label{fig:consistency}
\end{figure}

In order to check the consistency of our atmospheric parameters with previous studies,
    the results are compared with those given by \citet{silva2006}, \citet{Mishenina2006},
    \citet{Hekker2007}, \citet{Luck2007}, \citet{Takeda2008}, and \citet{Sato2010}.
All literatures above have determined the parameters of giants based on high resolution spectra.
There are totally 7 common stars in our sample with those literatures,
    as listed in table \ref{tab:commonstars}.
Figure \ref{fig:consistency} compares the effective temperatures, surface gravities, and metallicity
    of these stars.
Our derived temperatures are well consistent with others,
    with $\Delta T_\mathrm{eff}\sim4\pm65\,\mathrm{K}$ lower than their values.
While the surface gravity difference $\Delta\log g$ is $\sim0.14\pm0.33$ dex lower compared with others' results,
    and the metallicity difference $\Delta\mathrm{[Fe/H]}$ is $\sim0.09\pm0.07$ dex lower than those from literatures.

{\footnotesize
\begin{table}[htbp]
\caption{Common stars in this study and other literatures}
\label{tab:commonstars}
\begin{tabular}{c|l}
\hline
  HD   & Literature\\
\hline
 145457 & \cite{Sato2010}   \\
 161502 & \cite{Luck2007}   \\
 179799 & \cite{silva2006}  \\
 180314 & \cite{Sato2010}   \\
 185351 & \cite{Takeda2008} \\
        & \cite{Hekker2007} \\
 188993 & \cite{Luck2007}   \\
 196134 & \cite{Mishenina2006} \\
\hline
\end{tabular}
\end{table}
}

\section{Chemical abundances}

We use the atmospheric parameters listed in table \ref{tab:para_BV} to derive the [X/Fe] ratios
    of 14 elements (Al, Ba, Ca, Cr, Eu, K, La, Mg, Ni, Sc, Si, Ti, V and Y).
Those trends are plotted in figure \ref{fig:mg-si-ca-ti} to figure \ref{fig:y-ba-la-eu} (filled circles),
    together with the trends from Takeda et al. (2008, open circles) as a comparison.
The measured equivalent widths of the selected lines of our program stars are listed in table 9,
    which is only available in the electronic form.

\subsection{Error analysis}

The uncertainties of chemical abundances are estimated by
    changing the atmospheric parameters in a reasonable range.
Table \ref{tbl:error} shows the abundances differences due to deviations of the effective temperature of 100 K,
    the surface gravity of 0.15 dex,
    the iron abundance of 0.1 dex,
    and the microturbulent velocity of 0.1 $\mathrm{km\,s^{-1}}$.
For most of the chemical elements, the uncertainties are less than 0.1 dex, except of Ti and V.
The uncertainties of these two elements can be as high as 0.15 dex.

\begin{table}[htbp]
\caption{Estimated errors for abundances of a typical late-type giant HD109305,
    with $T_\mathrm{eff}=4663\,\mathrm{K}$, $\log g=2.44$, $\mathrm{[Fe/H]}=-0.30$,
    and $\xi_\mathrm{t}=1.77\,\mathrm{km\,s^{-1}}$, taken from table \ref{tab:para_BV}.}
\label{tbl:error}
\centering
\begin{tabular}{l|rrrrr}
\hline\hline
$\Delta [\mathrm{X/H}]$  & 
$\Delta T_\mathrm{eff}$  &
$\Delta\log g$           &
$\Delta\mathrm{[Fe/H]}$  &
$\Delta\xi_\mathrm{t}$   &
$\sigma_\mathrm{Total}$ \\
    &
\small ($+100\mathrm{K}$)    &
 ($+0.15$) &
 ($+0.1$)  &
 ($+0.1\,\mathrm{km\,s^{-1}}$) &
 \\

\hline
 $\Delta\mathrm{[FeI/H]}$ &  0.06  &  0.01  &  0.01  &-0.02  & 0.06 \\
 $\Delta\mathrm{[FeII/H]}$& -0.07  &  0.08  &  0.04  &-0.02  & 0.12 \\
 $\Delta\mathrm{[Mg/H]}$  &  0.05  & -0.01  &  0.01  &-0.01  & 0.05 \\
 $\Delta\mathrm{[Al/H]}$  &  0.07  & -0.01  &  0.00  &-0.02  & 0.07 \\
 $\Delta\mathrm{[Si/H]}$  & -0.02  &  0.03  &  0.02  &-0.01  & 0.04 \\
 $\Delta\mathrm{[Ca/H]}$  &  0.10  & -0.01  &  0.00  &-0.04  & 0.11 \\
 $\Delta\mathrm{[Sc/H]}$  & -0.01  &  0.06  &  0.03  &-0.03  & 0.07 \\
 $\Delta\mathrm{[Ti/H]}$  &  0.15  &  0.00  & -0.01  &-0.03  & 0.15 \\
 $\Delta\mathrm{[V/H]}$   &  0.15  &  0.00  &  0.00  &-0.01  & 0.15 \\
 $\Delta\mathrm{[Cr/H]}$  &  0.09  &  0.00  &  0.00  &-0.03  & 0.09 \\
 $\Delta\mathrm{[Ni/H]}$  &  0.05  &  0.03  &  0.02  &-0.02  & 0.06 \\
 $\Delta\mathrm{[Y/H]}$   & -0.01  &  0.07  &  0.04  & 0.00  & 0.08 \\
 $\Delta\mathrm{[Ba/H]}$  &  0.02  &  0.02  &  0.06  &-0.05  & 0.08 \\
 $\Delta\mathrm{[La/H]}$  &  0.02  &  0.07  &  0.04  &-0.01  & 0.08 \\
 $\Delta\mathrm{[Eu/H]}$  & -0.01  &  0.07  &  0.04  &-0.01  & 0.08 \\
 $\Delta\mathrm{[K/H]}$   &  0.10  & -0.04  &  0.01  &-0.05  & 0.12 \\
\hline
\end{tabular}
\end{table}

Another source of uncertainties is the error of equivalent widths caused by mixture of the intrinsic stellar lines
    and the $\mathrm{I}_2$ lines,
    since the spectra of our program stars are $\mathrm{I}_2$-superposed for precise radial velocity measurements.
To check the uncertainties caused by iodine lines, we use the pure stellar spectra of HD 145457 taken in OAO
    to measure the equivalent widths, and determine the chemical abundances.

\begin{figure}
  \begin{center}
     \FigureFile(70mm, 70mm){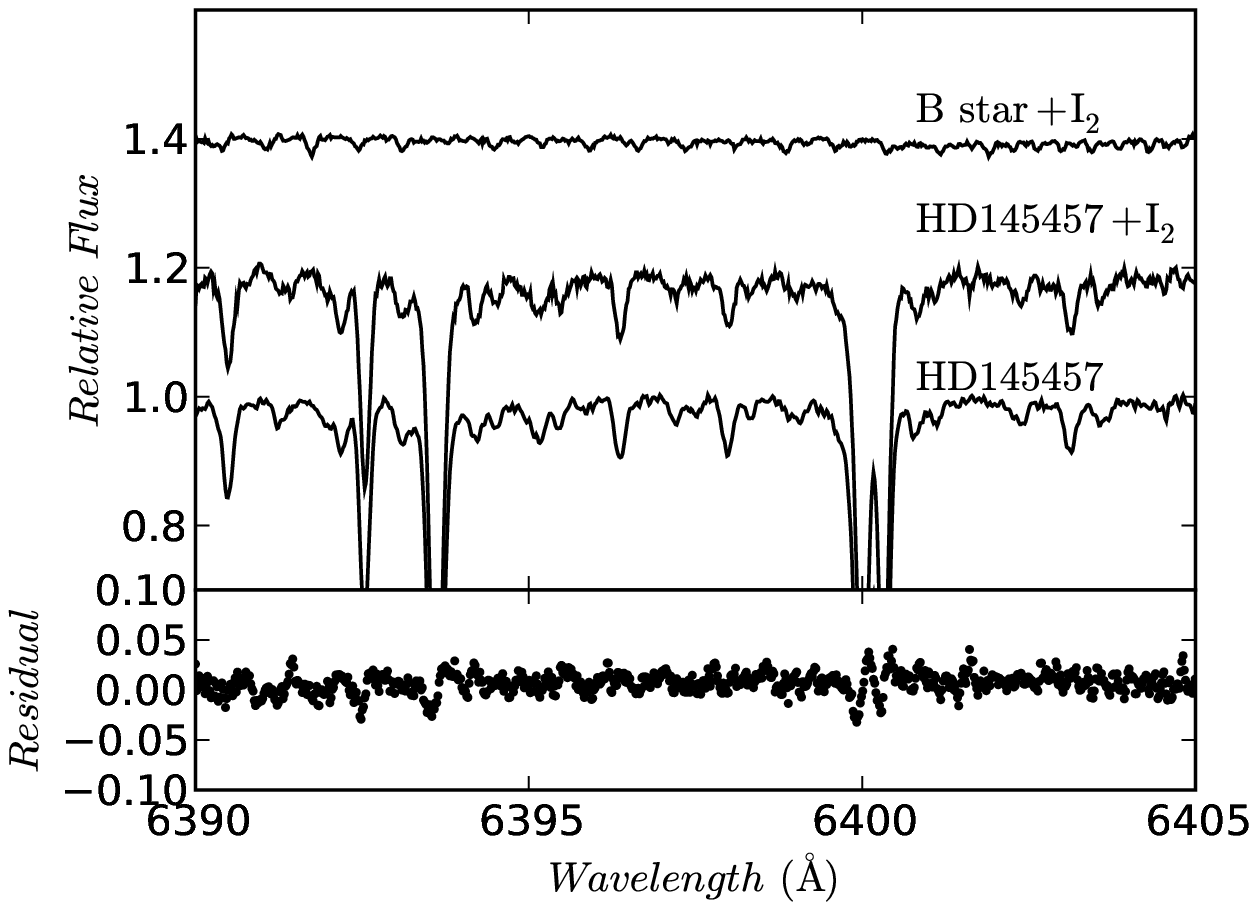}
     \FigureFile(70mm, 70mm){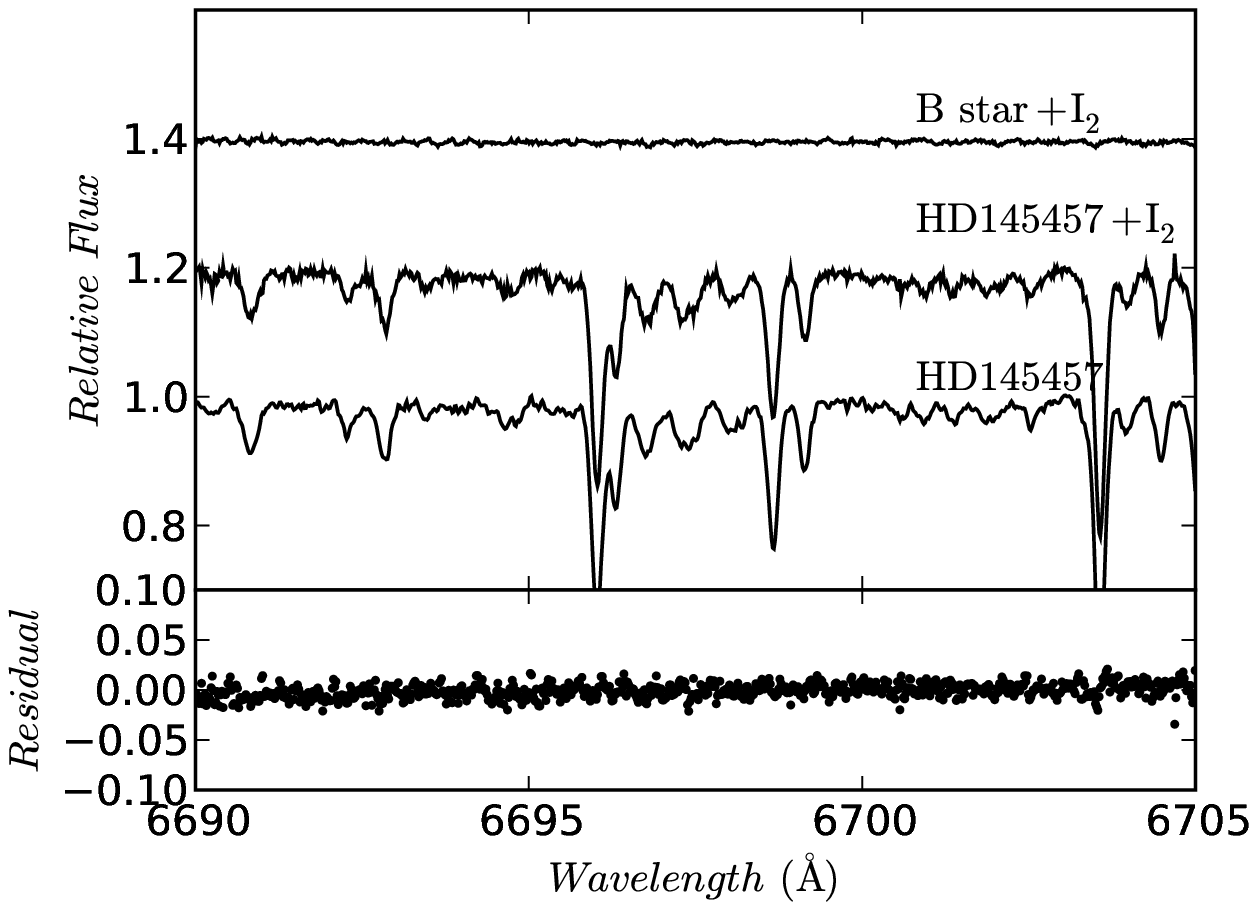}
  \end{center}
  \caption{
Comparison of the $\mathrm{I_2}$-superposed spectrum (taken in Subaru) and the pure spectrum (taken in OAO)
    for HD 145457 around $6400\,\mathrm{\AA}$ and $6700\,\mathrm{\AA}$.
The calculated RMS residuals are 1.2\% around $6400\,\mathrm{\AA}$ and 0.7\% around $6700\,\mathrm{\AA}$.
Two portions of $\mathrm{I_2}$-superposed spectrum for
    a fast rotation, B-type star (HR 5685, taken in Subaru) are also plotted in the upper panel.
}
  \label{fig:spectra_seg}
\end{figure}

In Doppler surveys, spectra of fast-rotation B type stars are often obtained with $\mathrm{I_2}$ cell
   to deconvolute instrumental profiles (IPs, e.g. \cite{Butler1996}).
The spectra of such stars are featureless, and thus provide a simple way to check
   how the stellar spectra are contaminated by $\mathrm{I_2}$ absorption lines
   at different wavelengths in this study.
The $\mathrm{I_2}$-superposed spetra of a B star (HR 5685, B8V), $\mathrm{I_2}$-superposed spetra of HD 145457,
    and the pure spectra of HD 145457 around $6400\,\mathrm{\AA}$ and $6700\,\mathrm{\AA}$ are plotted
    in figure \ref{fig:spectra_seg}.
Around $6400\,\mathrm{\AA}$, the integrated equivalent width of $\mathrm{I_2}$ lines is about $6\,\mathrm{m\AA/\AA}$,
    which causes the uncertainty of equivalent width of about $3\sim4\,\mathrm{m\AA}$ for a typical spectral line
    in our $\mathrm{I_2}$-superposed spetrum in this region.
While around $6700\,\mathrm{\AA}$, the integrated equivalent width of $\mathrm{I_2}$ lines is about $4\,\mathrm{m\AA/\AA}$,
    30\% smaller than that around $6400\,\mathrm{\AA}$.
The uncertainty of equivalent width caused by $\mathrm{I_2}$ lines is estimated to be $2\,\mathrm{m\AA}$.

\begin{figure}
  \begin{center}
     \FigureFile(70mm, 70mm){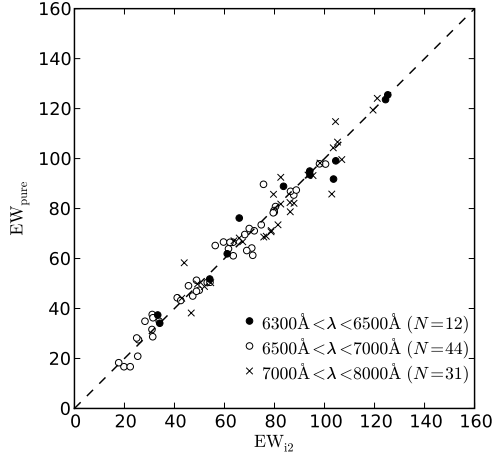}
  \end{center}
  \caption{Comparison of equivalent widths (EWs) measured using $\mathrm{I}_2$-superposed spectra taken in Subaru (X axis)
            and pure spectra taken in OAO (Y axis). The spectral lines are separated into three groups according to
            their wavelengths. The filled circles are 12 lines with wavelengths $\lambda<6500\,\mathrm{\AA}$.
            The open circles are 44 lines with wavelengths $6500\,\mathrm{\AA}<\lambda<7000\,\mathrm{\AA}$.
            The crosses are 31 lines with wavelengths $\lambda>7000\,\mathrm{\AA}$.
            The mean differences $\langle EW_\mathrm{I2} - EW_\mathrm{pure}\rangle$ are
    $-0.05 \pm 5.22 \,\mathrm{m\AA}$,
    $-0.33 \pm 4.33 \,\mathrm{m\AA}$, and
    $+1.48 \pm 6.38 \,\mathrm{m\AA}$, respectively.
}
  \label{fig:comp_ew}
\end{figure}

Furthermore, we measure the equivalent widths using the $\mathrm{I}_2$-superposed spctrum
    and pure spectrum of HD 145457.
The spectral lines are separated into three groups, with wavelengths $\lambda<6500\,\mathrm{\AA}$,
    $6500\,\mathrm{\AA}<\lambda<7000\,\mathrm{\AA}$, and $\lambda>7000\,\mathrm{\AA}$, respectively.
The results are compared in figure \ref{fig:comp_ew}.
We find that the mean differences $\langle EW_\mathrm{I2} - EW_\mathrm{pure}\rangle$ of three groups are
    $-0.05 \pm 5.22 \,\mathrm{m\AA}$,
    $-0.33 \pm 4.33 \,\mathrm{m\AA}$, and
    $+1.48 \pm 6.38 \,\mathrm{m\AA}$, which infers our EW values are not significantly influenced
    by $\mathrm{I}_2$ absorption lines.
The chemical abundaces determined by pure stellar spectra are compared with our results using Subaru spectra
    in table \ref{tbl:error_i2}.
It is shown that the uncertainties of chemical abundances caused by mixture of $\mathrm{I}_2$ lines are range from
    $0.02 \sim 0.09$ dex, which are smaller than the uncertainties caused by errors of atmospheric parameters.
Therefore, we concluded that the weak $\mathrm{I}_2$ absorption lines in the red part of our spectra nearly do not
    affect the chemical abundances measurements.

\begin{table}[htbp]
\caption{Comparison of chemical abundances of HD 145457 determined by pure stellar specta taken in OAO,
        and $\mathrm{I}_2$ superposed spectra taken in Subaru.}
\label{tbl:error_i2}
\centering
\begin{tabular}{l|c}
\hline\hline
$\Delta [\mathrm{X/H}]$ & $\mathrm{[X/H]_{i2} - [X/H]_{pure}}$ \\
\hline

$\Delta\mathrm{[FeI/H]}$ & 0.02 \\
$\Delta\mathrm{[FeII/H]}$& 0.04 \\
$\Delta\mathrm{[Mg/H]}$  & 0.04 \\
$\Delta\mathrm{[Al/H]}$  & 0.03 \\
$\Delta\mathrm{[Si/H]}$  & 0.04 \\
$\Delta\mathrm{[Ca/H]}$  & 0.09 \\
$\Delta\mathrm{[Sc/H]}$  & 0.06 \\
$\Delta\mathrm{[Ti/H]}$  & 0.04 \\
$\Delta\mathrm{[V/H]}$   & 0.02 \\
$\Delta\mathrm{[Cr/H]}$  & 0.07 \\
$\Delta\mathrm{[Ni/H]}$  & 0.07 \\
$\Delta\mathrm{[Y/H]}$   & 0.02 \\
$\Delta\mathrm{[Ba/H]}$  & 0.02 \\
$\Delta\mathrm{[La/H]}$  & 0.03 \\
$\Delta\mathrm{[Eu/H]}$  & 0.02 \\
$\Delta\mathrm{[K/H]}$   &  --  \\

\hline
\end{tabular}
\end{table}

\subsection{$\alpha$-elements (Mg, Si, Ca, Ti)}

\begin{figure}
  \begin{center}
     \FigureFile(70mm,120mm){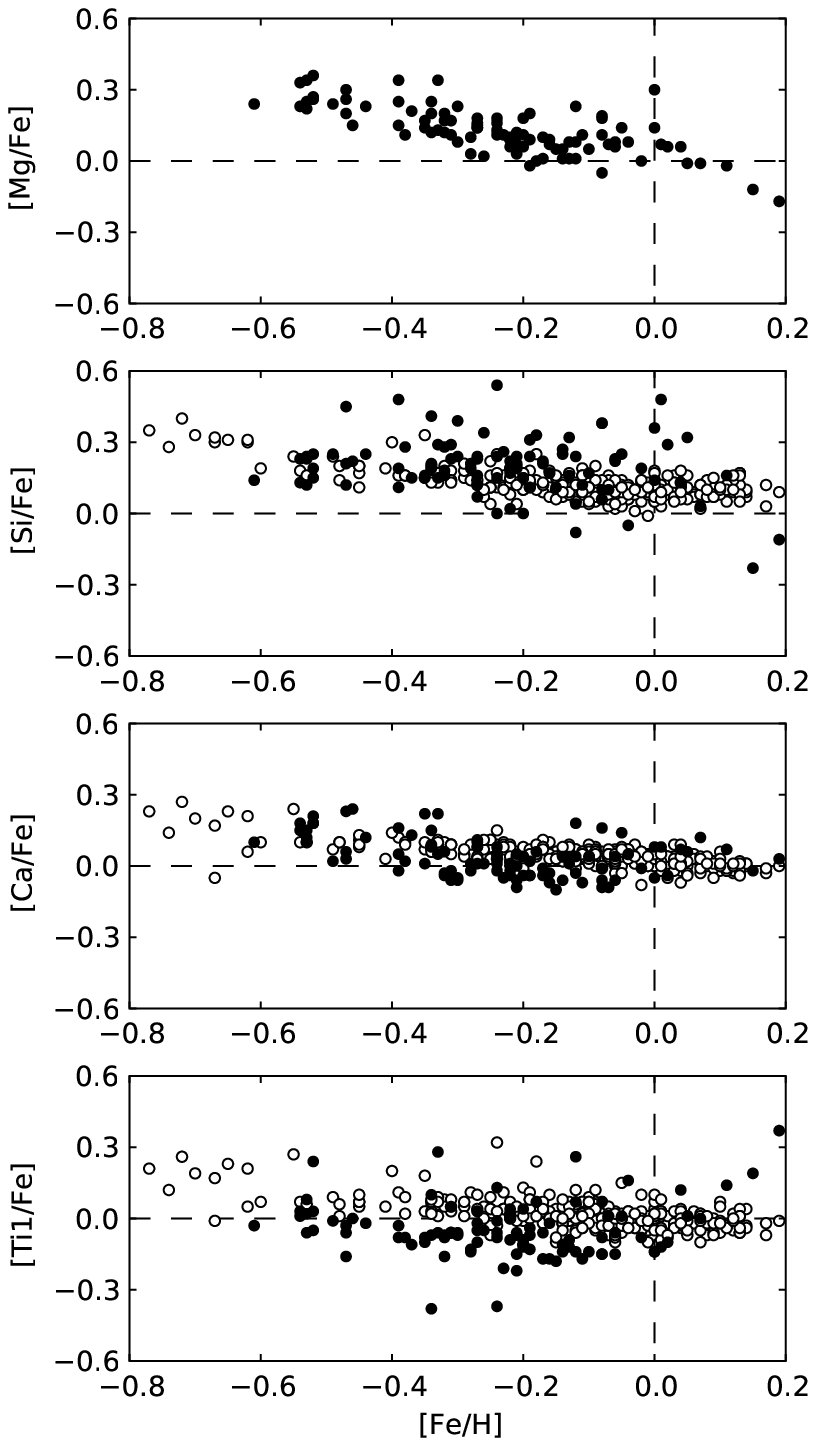}
  \end{center}
  \caption{Abundance ratio [X/Fe] for four $\alpha$ elements (Mg, Si, Ca, and Ti I) against [Fe/H] (filled circles),
    together with the results from Takeda et al. (2008, open circles).}
  \label{fig:mg-si-ca-ti}
\end{figure}

The $\alpha$ elements are mainly produced in SNe II nucleosynthesis \citep{WW1995} and show
    enhancement in metal-poor stars.
In figure \ref{fig:mg-si-ca-ti}, we plot [X/Fe] ratios of Mg, Si, Ca, and Ti I versus [Fe/H],
    together with the trends take from \citet{Takeda2008}.
Except for the trend of [TiI/Fe] versus [Fe/H] with a large scatter, three of the other $\alpha$ elements exhibit
    enhancement towards lower metallicity.
However, their decreasing trends with increasing metallicity are slightly different.
Both the [Mg/Fe] and [Ca/Fe] show a turn off trend towards a flat pattern at $\mathrm{[Fe/H]}\sim-0.2$,
    which is in good agreement with previous studies on both giants (\cite{Takeda2008}, \cite{Liu07})
    and dwarfs (\cite{Chen00}), while the [Mg/Fe] turns to decrease at $\mathrm{[Fe/H]}>0$.
Despite an increasing trend with decreasing metallicity for $\mathrm{[Fe/H]}<-0.4$,
    [Si/Fe] shows a larger scatter for higher metallicity, leaving the flat pattern
    shown in \citet{Takeda2008} (open circles in figure \ref{fig:mg-si-ca-ti}) unclear.
Oxygen is another important $\alpha$ element for giants.
But in our spectra, the [O I] forbidden lines at 6300\ \AA\ are not covered by the CCD,
    and the lines at 6363\,\AA\ are very weak and closed to the edge.
To make sure our results are reliable, the oxygen abundance is not included in this study.

\subsection{Odd-Z light elements (Al, K, Sc)}

\begin{figure}
  \begin{center}
     \FigureFile(70mm,120mm){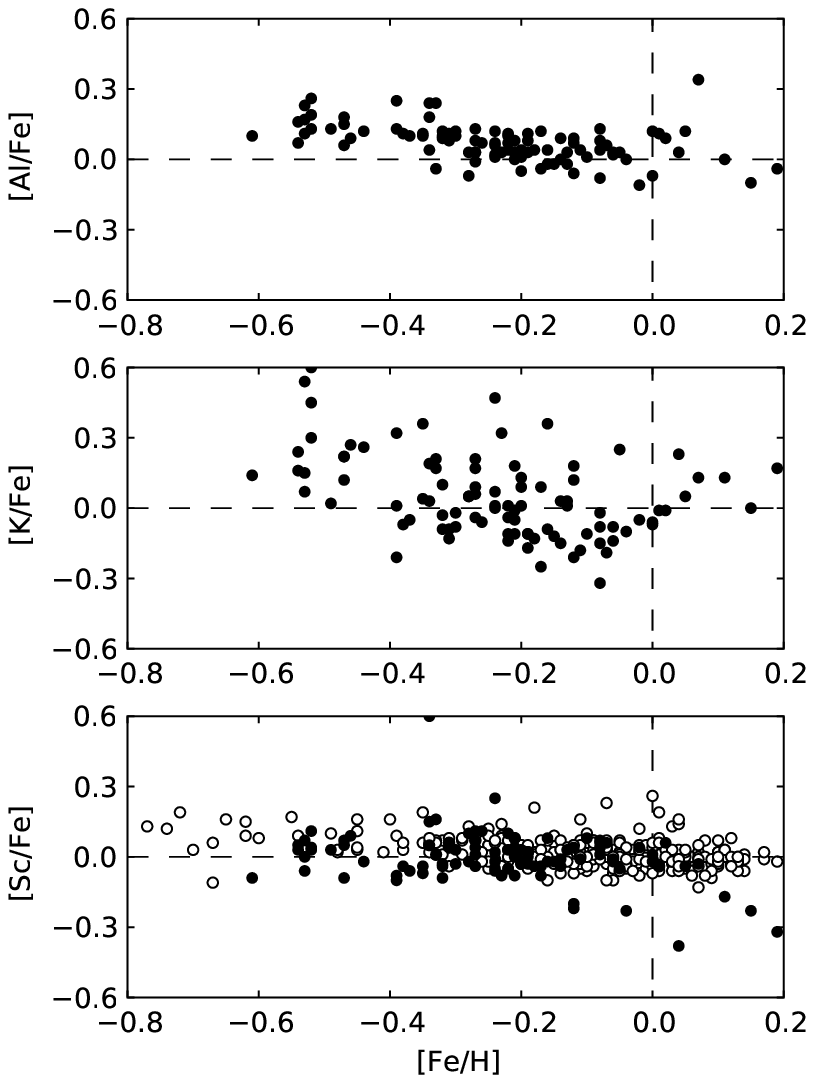}
  \end{center}
  \caption{Abundance trends of [X/Fe] against [Fe/H] for three odd-z elements (Al, K, and Sc, filled circles).
           Results of [Sc/Fe] from \citet{Takeda2008} are plotted as open circles.}
  \label{fig:al-k-sc}
\end{figure}

As shown in figure \ref{fig:al-k-sc},
    [Al/Fe] shows a decreasing trend with increasing metallicity for $\mathrm{[Fe/H]}<-0.2$,
    and a solar pattern towards higher metallicity.
Our results are consistent with \citet{Liu07}.
\citet{Chen00} found a rather steep upturn of [Al/Fe] at $\mathrm{[Fe/H]}=-0.2$ in a sample of dwarfs.
However, in our results the upturn is rather smooth.
[K/Fe] exhibits a larger dispersion, because only very few number of lines are available.
The trend of [Sc/Fe] is similar with that of [Ti/Fe], with both of them showing a flat pattern
    when data with larger dispersion are excluded.
This result is also consistent with \citet{Takeda2008}.

\subsection{Iron-peak elements (V, Cr, Ni)}

\begin{figure}
  \begin{center}
     \FigureFile(70mm,120mm){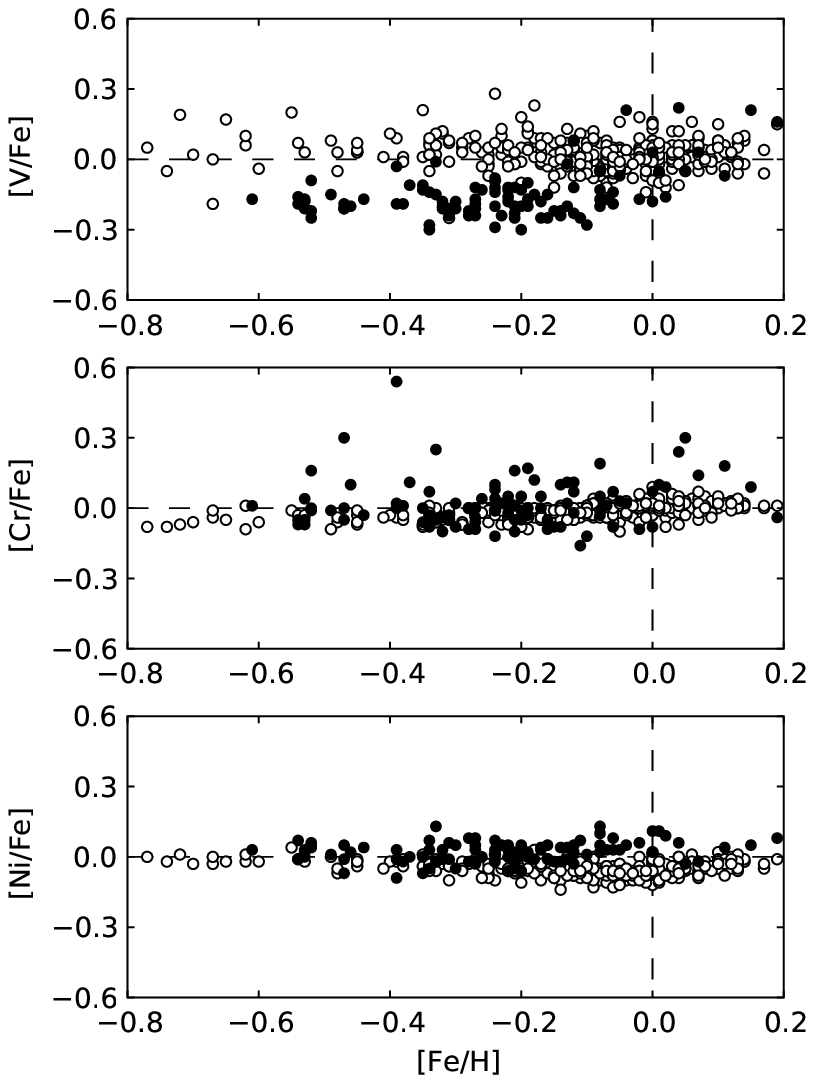}
  \end{center}
  \caption{Abundance trends of [X/Fe] against [Fe/H] for three iron-peak elements (V, Cr, and Ni, filled circles).
           Results from \citet{Takeda2008} are plotted as open circles.}
  \label{fig:v-cr-ni}
\end{figure}

The iron-peak elements are believed to be mainly produced in SNe Ia,
    and should have the same abundance patterns with iron.
In figure \ref{fig:v-cr-ni}, we plot [X/Fe] versus [Fe/H] for three iron-peak elements (V, Cr, and Ni).
[V/Fe] shows an increasing trend with increasing metallicity, which is contrast to Takeda's results
    (plotted as open circles in the same figure).
The hyper fine structure (HFS) effect has been considered in our work.
The adopted HFS data is taken from Kurucz \footnote{http://kurucz.harvard.edu/linelists.html}.
\citet{Liu07} suggests the HFS effect can lead to a correction as large as 0.5 dex
    for vanadium abundance in giants.
However, we find that the corrections due to introduction of HFS effect are between -0.002 dex and 0.012 dex,
    and can be neglected (see figure 12).
The discrepancy is due to the difference of HFS effect from line to line,
    and the vandium lines we used in this work are different from those in \citet{Liu07}.

\begin{figure}
  \begin{center}
     \FigureFile(70mm,120mm){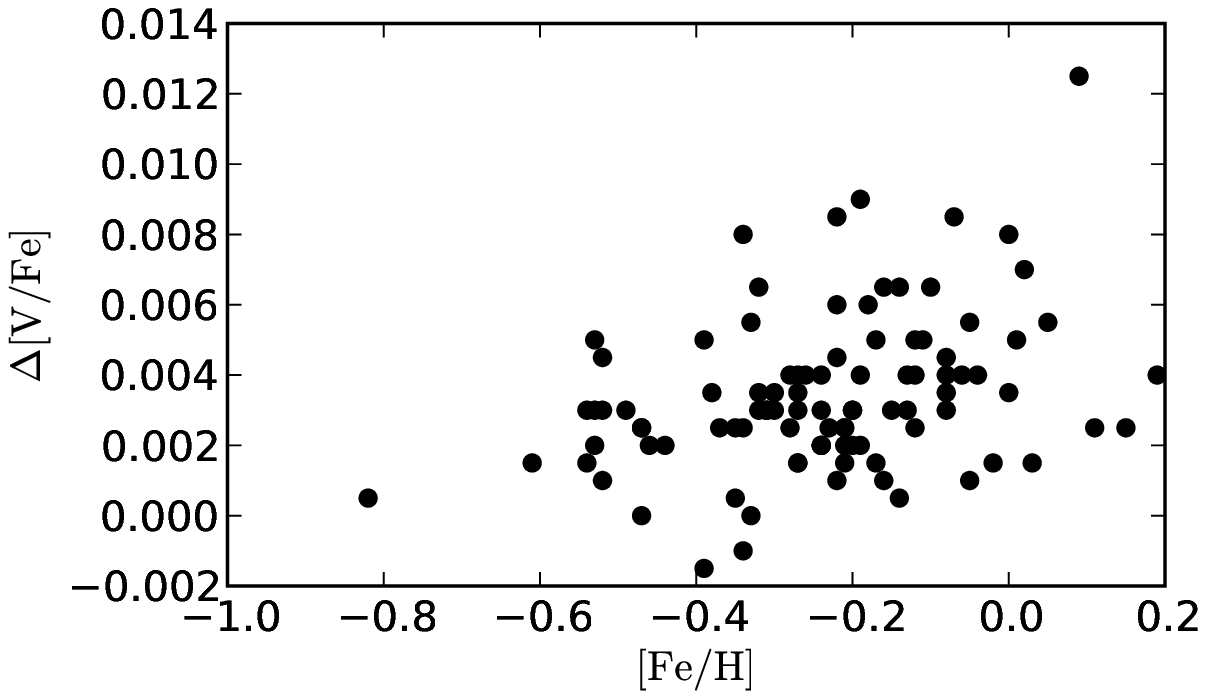}
  \end{center}
  \caption{The difference in vandium abundances obtained by neglecting HFS minus those by considering HFS versus
           [Fe/H]. $\Delta\mathrm{[V/Fe]}$ on Y axis represents $\log\mathrm{V_{non-HFS}}-\log\mathrm{V_{HFS}}$.
           }
  \label{fig:v-hfs}
\end{figure}

The [Cr/Fe] trend in our sample is consistent with Takeda's,
    except for a few stars with higher [Cr/Fe] values.
For these stars, the relation of [Ni/Fe] vs. [Fe/H] keeps the solar pattern in the range of $-0.8<\mathrm{[Fe/H]}<-0.1$,
    while shows overabundance for higher metallicity and hence systematically higher than Takeda's result.
The possible upturn at $\mathrm{[Fe/H]}\sim0.0$ has also been found in giants (e.g. \citet{Liu07}) and dwarfs (e.g. \cite{Edvardsson1993}).

\subsection{Heavy elements (Y, Ba, La, Eu)}

\begin{figure}
  \begin{center}
     \FigureFile(70mm,120mm){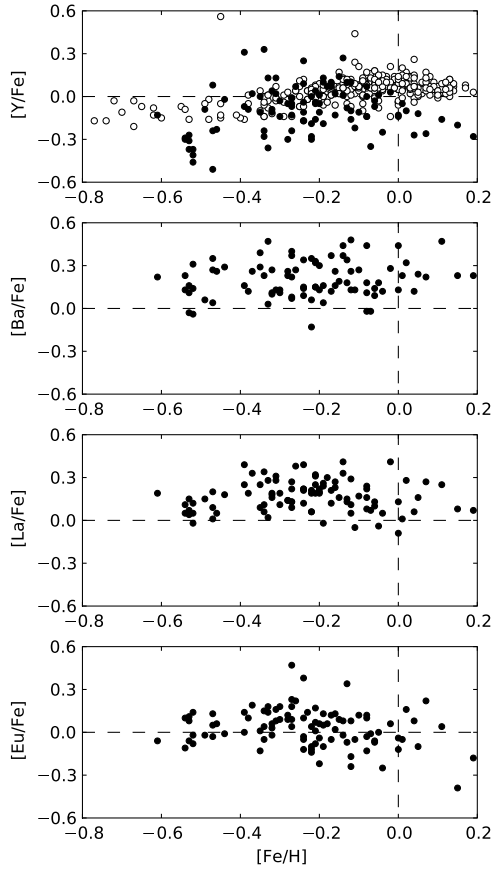}
  \end{center}
  \caption{Abundance trends of [X/Fe] against [Fe/H] for four heavy elements (Y, Ba, La, and Eu).
           (filled circles: this work; open circles: \cite{Takeda2008}'s work)}
  \label{fig:y-ba-la-eu}
\end{figure}

Those elements heavier than iron are mainly produced via the neutron capture process.
Two major mechanisms are generally considered: the r-process and s-process,
    corresponding to neutron rich and neutron poor environment, respectively.
In our analysis, one r-process element (Eu), one light s-process element (Y) and two
    heavy s-process elements (Ba, La) are included.
Their abundances trend against [Fe/H] are plotted in figure \ref{fig:y-ba-la-eu}.

\begin{figure}
  \begin{center}
     \FigureFile(70mm,120mm){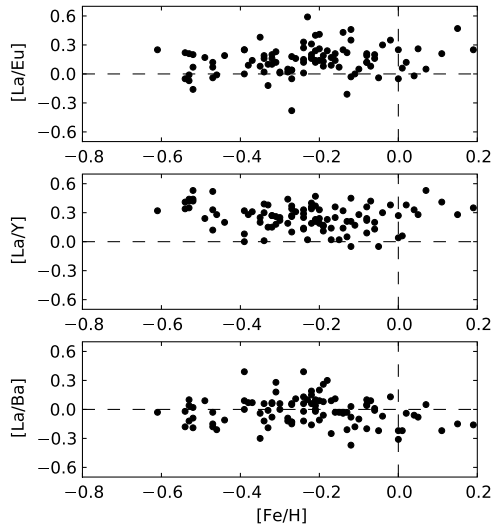}
  \end{center}
  \caption{[La/Eu], [La/Y], [La/Ba] as a function of [Fe/H] (from top to bottom).}
  \label{fig:la}
\end{figure}

Despite a large scatter as often observed in field stars,
    the ratio of [La/Eu], [La/Y], and [La/Ba] may provide important information of the Galactic
    chemical evolution history.
As a typical r-process element, europium is mainly produced in core-collapse supernovae (SNe) with masses $8\,M_\odot<M<10\,M_\odot$
    and the pattern still remains uncertain.
Previous work \citep{Simmerer2004} of La and Eu revealed the s-process may be active
    as early as [Fe/H]=-2.6.
In figure \ref{fig:la}, [La/Eu], [La/Y], and [La/Ba] are plotted as functions of [Fe/H] (from top to bottom).
From our plot, a decline of [La/Eu], although not very clearly,
    can be seen in lower metallicity down to -0.7.
In the convection zone of low mass AGB stars where the slow neutron capture reactions
    take place, the neutron flux per seed nuclei is inversely proportional to the density
    of iron-peak nucleus.
The metal-poor stars produce light s-process elements (such as Y) more efficiently than
    the heavy s-process elements (such as Ba, La).
The [La/Y] versus [Fe/H] in the middle panel of figure \ref{fig:la} shows an marginal increasing trend
    with decreasing metallicity, consistent with the theoretical prediction (e.g. \cite{Busso1999}).
[La/Ba] shows a flat trend with [Fe/H] in the bottom panel of figure \ref{fig:la}.
This is consistent with the theoretical prediction that they are produced through the same producing mechanism.

\section{Kinetics Parameters}
\begin{figure}
  \begin{center}
     \FigureFile(70mm,70mm){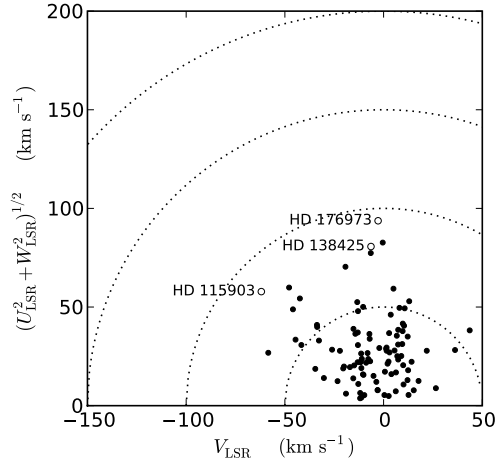}
  \end{center}
  \caption{Toomre diagram of program stars. Thin disk stars ($\mathrm{TD/D}<1.0$)
           and thick disk stars ($\mathrm{TD/D}>1.0$) are plotted as filled and open circles,
           respectively (See discussion in Sect. 5). Dotted lines indicate constant total velocities
           $|v|_\mathrm{LSR}=\sqrt{U_\mathrm{LSR}^2+V_\mathrm{LSR}^2+W_\mathrm{LSR}^2}$
           in steps of 50 $\mathrm{km\,s^{-1}}$.}
  \label{fig:toomre}
\end{figure}

We derived the kinetics parameters ($U_\mathrm{LSR}$, $V_\mathrm{LSR}$, $W_\mathrm{LSR}$)
    based on the method given by \citet{Johnson1987}.
The radial velocities ($RV_\mathrm{helio}$) are measured based on our spectra,
    and corrected to heliocentric velocities.
Positions, parallaxes, and proper motions are taken from {\em Hipparcos} data.
By adopting a solar motion of
    $(U, V, W)_\odot = (-10.00\pm0.36, +5.25\pm0.62, +7.17\pm0.38)$
    \footnote{In this work the Galactic velocity component $U$ is defined to be positive towards Galactic anticenter.}
    in $\mathrm{km\,s^{-1}}$, given by \citet{Dehnen1998},
    the Galactic velocity components U, V, and W of our program stars are corrected to local standard of rest (LSR).
It is hard to clarify whether a star in solar neighborhood belongs to the thin disk or the thick disk.
We apply the kinematic method proposed by \citet{Bensby2003} to calculate the relative probability for
    the thick-disk-to-thin-disk (TD/D) membership.
In \citet{Bensby2003}, thin disk and thick disk stars are thought to follow Gaussian distributions
    in Galactic velocity space ($U$, $V$, $W$).
By assuming different asymmetric velocity drifts and dispersions, the relative probability of
    belonging to different populations can be calculated.
For instance, a star with $\mathrm{TD/D}=10$ means it is 10 times more likely
    to belong to the thick disk than to the thin disk.
In our sample, one star (HD 115903) with $\mathrm{TD/D}>10.0$ and two stars (HD 138425 and HD 176973)
    with $1.0<\mathrm{TD/D}<10.0$ are more likely to be a thick disk star rather than a thin disk star.
Those stars are plotted as open circles on Toomre diagram in figure \ref{fig:toomre}, and listed with a mark `TD'
    in the last column of table \ref{tab:kinetics}, as well as their kinetics parameters mentioned above.

{\footnotesize
\begin{table}[htbp]
\caption{Kinetics parameters of program stars
}
\label{tab:kinetics}
\begin{tabular}{c|rrrrl}
\hline
  HD   & $RV_\mathrm{helio}$  & $U_\mathrm{LSR}$ & $V_\mathrm{LSR}$ & $W_\mathrm{LSR}$ & TD/D \\
 & ($\mathrm{km\,s^{-1}}$) & ($\mathrm{km\,s^{-1}}$) & ($\mathrm{km\,s^{-1}}$) & ($\mathrm{km\,s^{-1}}$) & \\
\hline
 100055 &   6.69 $\pm$ 0.37 &   2.9 $\pm$  1.3 & -15.4 $\pm $ 2.6 &  13.5 $\pm$  0.7 &  0.01 \\
 101853 &   2.81 $\pm$ 0.36 & -30.9 $\pm$  3.4 &   6.7 $\pm $ 0.7 &  17.3 $\pm$  1.3 &  0.01 \\
 103690 & -19.28 $\pm$ 0.35 & -58.6 $\pm$  3.8 &   5.0 $\pm $ 1.1 &   8.9 $\pm$  1.7 &  0.02 \\
 105475 &   0.35 $\pm$ 0.35 & -80.9 $\pm$ 11.8 &  -0.4 $\pm $ 1.3 &  16.9 $\pm$  1.7 &  0.06 \\
  ...   &   ...           & ...            &  ...           &  ...           &  ...  \\
\hline
\end{tabular}
\end{table}

}

\section{Conclusion}
In this work we determine the atmospheric parameters,
    kinematic properties and chemical abundances of 99 late-type giants
    based on high resolution spectra obtained in Subaru Planet Search Program,
    covering the metallicity range $-0.8<\mathrm{[Fe/H]}<+0.2$.
We get two sets of stellar parameters derived from different methods.
The photospheric chemical abundances of 15 elements, including
    four $\alpha$-elements (Mg, Si, Ca, Ti),
    three odd-Z light elements (Al, K, Sc),
    four iron peak elements (V, Cr, Fe, Ni),
    and four neutron-capture elements (Y, Ba, La, Eu)
    are determined.
The kinematic parameters are calculated, and most of our sample are thin disk stars.
From the results of abundances, we conclud that
    the abundance ratios [Mg/Fe] and [Ca/Fe] show an increasing trend with decreasing metallicty for $-0.8<\mathrm{[Fe/H]}\lesssim-0.2$,
    while flatten out towards higher metallicity.
    A decreasing trend of [Mg/Fe] with increasing metallicity is detected for giants at $\mathrm{[Fe/H]>0}$.
The upturn of [Al/Fe] versus [Fe/H] at $\mathrm{[Fe/H]}=-0.2$ in giants is not as steep as that in dwarfs (e.g. \cite{Chen00}).
The [Ni/Fe] versus [Fe/H] shows an upturn at [Fe/H]=0.0 and an overabundance towards higher metallicity.
The [La/Y] ratio shows an increasing trend with decreasing metallicity,
    which is consistent with the prediction of AGB nucleosynthesis.

\section*{Acknowledgements}
WL is grateful of Prof. Chen Yuqin and Dr. Li Haining for valuable comments and discussions.
This work is supported by the National Natural Science Foundation of China
    under grant number 10821061 and 10803010.

\end{document}

%% file: table.teff_comp.dat
 101853& 4841& 4834& 4853& 120651& 4916& 4885& 4855& 159797& 4811& 4832& 4785\\
 103690& 4694& 4730& 4743& 121303& 4888& 4994& 4854& 160823& 4891& 5076& 5100\\
 105475& 4787& 4764& 4745& 121883& 4876& 4904& 4905& 161502& 4936& 4975& 4951\\
 109545& 5001& 5006& 5065& 125113& 4873& 4896& 4930& 174621& 5168& 5205& 5168\\
 109894& 4830& 4848& 4849& 137463& 5536& 5590& 5225& 176973& 4812& 4780& 4852\\
 111381& 4867& 4876& 4913& 138425& 5023& 4879& 5026& 181878& 5164& 5223& 5260\\
 113021& 4814& 4844& 4871& 149522& 4883& 4922& 4972& 183585& 5272& 5246& 5185\\
 114877& 4882& 4876& 4924& 151879& 5018& 5042& 5085& 188993& 5728& 5703& 5797\\
 114988& 5269& 5241& 5136& 155231& 4857& 4923& 4902& 194579& 5006& 5008& 5097\\
 117877& 4908& 4852& 4940& 157261& 4954& 4918& 4911& 195194& 4870& 4904& 4948\\
 118670& 4849& 4841& 4817& 159466& 4816& 4872& 5132\\